\documentclass[acmsmall]{acmart}

\AtBeginDocument{%
  }

\usepackage{xcolor}
\definecolor{darkgreen}{rgb}{0.0, 0.4, 0.3}

\setcopyright{acmlicensed}
\acmJournal{PACMHCI}
\acmYear{2024} \acmVolume{8} \acmNumber{CSCW2} \acmArticle{374} \acmMonth{11}\acmDOI{10.1145/3686913}

\begin{document}

\title{Characterizing the Structure of Online Conversations Across Reddit}

\author{Yulin Yu}
\email{yulinyu@umich.edu}
\orcid{0000-0003-4743-8360}
\affiliation{%
  \institution{University of Michigan, School of Information}
  \city{Ann Arbor}
  \state{Michigan}
  \country{USA}
}

\author{Julie Jiang}
\affiliation{%
  \institution{University of Southern California, Information Sciences Institute}
  \city{Los Angeles}
  \state{California}
  \country{USA}}
\email{juliej@isi.edu}
\orcid{0000-0003-4260-282X}

\author{Paramveer S. Dhillon}
\affiliation{%
  \institution{University of Michigan, School of Information}
  \city{Ann Arbor}
  \state{Michigan}
  \country{USA}
}
\email{dhillonp@umich.edu}
\orcid{0000-0002-0994-9488}

\renewcommand{\shortauthors}{Yulin Yu, Julie Jiang, and Paramveer S. Dhillon}
\renewcommand{\shortauthors}{Yu et al.}

\begin{abstract}
The proliferation of social media platforms has afforded social scientists unprecedented access to vast troves of data on human interactions, facilitating the study of online behavior at an unparalleled scale. These platforms typically structure conversations as threads, forming tree-like structures known as ``discussion trees.'' This paper examines the structural properties of online discussions on Reddit by analyzing both global (community-level) and local (post-level) attributes of these discussion trees. We conduct a comprehensive statistical analysis of a year's worth of Reddit data, encompassing a quarter of a million posts and several million comments. Our primary objective is to disentangle the relative impacts of global and local properties and evaluate how specific features shape discussion tree structures. The results reveal that both local and global features contribute significantly to explaining structural variation in discussion trees. However, local features, such as post content and sentiment, collectively have a greater impact, accounting for a larger proportion of variation in the width, depth, and size of discussion trees. Our analysis also uncovers considerable heterogeneity in the impact of various features on discussion structures. Notably, certain global features play crucial roles in determining specific discussion tree properties. These features include the subreddit's topic, age, popularity, and content redundancy. For instance, posts in subreddits focused on politics, sports, and current events tend to generate deeper and wider discussion trees. This research enhances our understanding of online conversation dynamics and offers valuable insights for both content creators and platform designers. By elucidating the factors that shape online discussions, our work contributes to ongoing efforts to improve the quality and effectiveness of digital discourse.
\end{abstract}




\received{July 2023}
\received[revised]{April 2024}
\received[accepted]{July 2024}

\maketitle

\section{Introduction \& Background}

The rapid expansion of the Internet in recent years has significantly reduced barriers to computer-mediated communication, ushering in a new era of online interaction. Today, online discussions via social media platforms\textemdash such as Reddit, Facebook, Instagram, and X\textemdash have become the dominant medium for conversations about events, ideas, and knowledge~\cite{anderson2014social,leonardi2014social,kittur2007he,diakopoulos2011towards,chancellor2018norms}. These platforms have transformed the communication landscape, allowing users to engage in dialogues that transcend geographical boundaries. The exponential growth in online discussion participants has led to a corresponding increase in the volume and diversity of shared content. This surge in content creation and sharing necessitates effective content moderation strategies to manage and filter information~\cite{berry2017discussion, heatherly2017filtering,rhue2014digital,chandrasekharan2017you,cheng2017anyone,kriplean2012you}, which is crucial for maintaining a digital space free from misinformation~\cite{del2016spreading}, online hate speech~\cite{chetty2018hate}, and other undesirable online behaviors.

As these online discussions have become increasingly prevalent and complex, researchers have recognized their significant social impact and have begun to investigate how the structural properties of these conversations influence discourse quality~\cite{kumar2010dynamics,backstrom2013characterizing,gomez2013likelihood,gomez2008statistical,gomez2011modeling,aragon2017thread,aragon2017generative,kaltenbrunner2007description,zhang2018characterizing}. Many studies have focused on predicting discussion outcomes based solely on size metrics, such as reply count or overall popularity~\cite{zhang2018conversations}. While this approach provides valuable insights, it overlooks the nuanced nature of discussions with similar sizes. Discussions can develop in two distinct ways: they may become {\it deep}, with users engaging in extended exchanges on specific topics, or {\it wide}, where multiple users offer diverse perspectives in response to the original post. This distinction is crucial because it reflects different patterns of user engagement and information exchange, which can have varying impacts on the quality and outcomes of online discourse. Understanding these structural differences is essential for developing more effective moderation strategies, improving platform design, and fostering healthier online communities.

\begin{figure}
\centering
% Note: text {\it Width} is the {\it Width} of both columns
\includegraphics[width=0.90\textwidth]{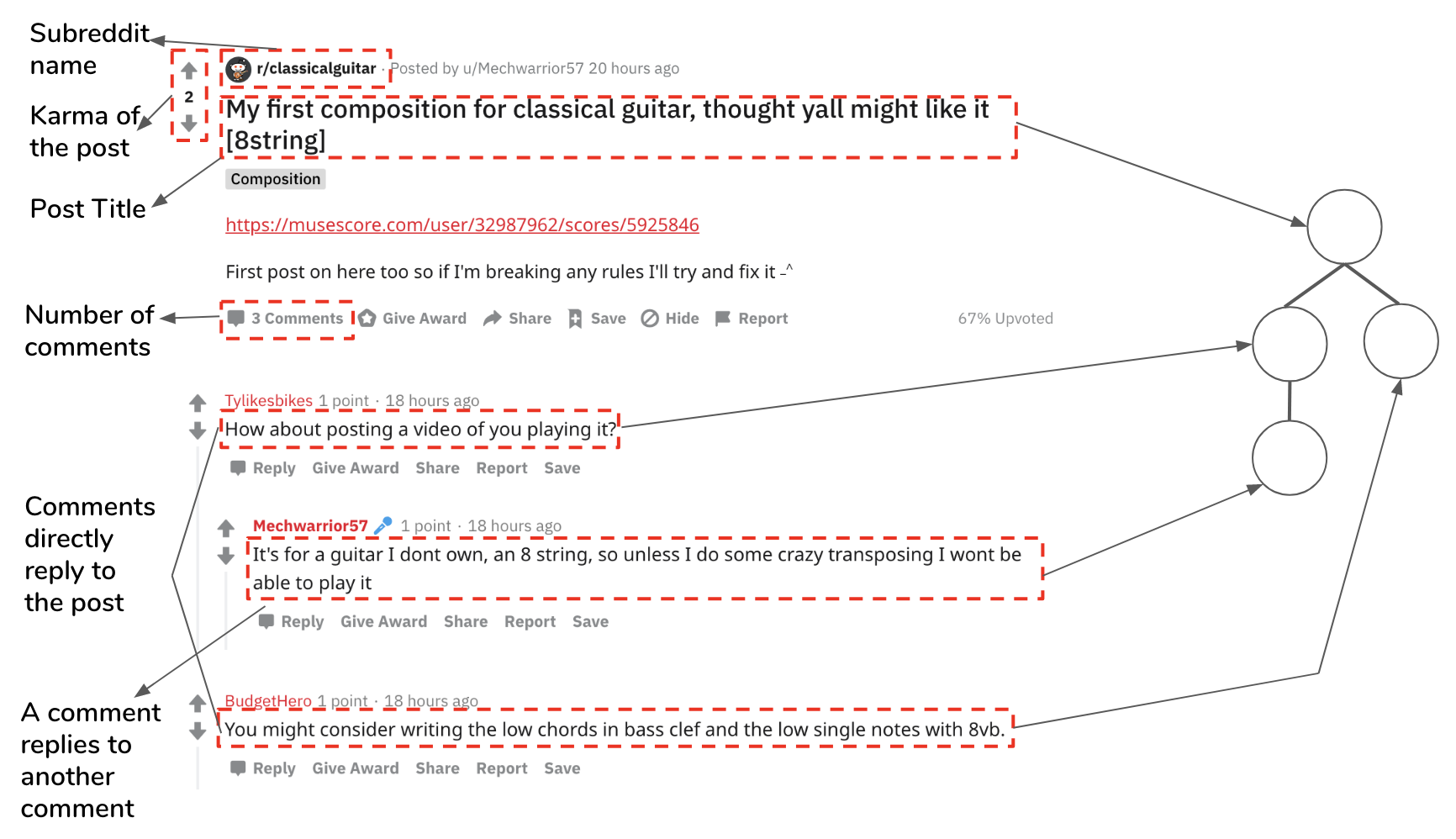}
\caption{Illustration of a Reddit post and its comment tree. In this paper, we seek to explain the structure of these discussion/comment trees using {\it local} and {\it global} properties of the post title. {\it Note:} 1) A given post on Reddit can be upvoted or downvoted. The \textit{karma} of a post is the difference between the number of upvotes and downvotes. 2) A Reddit post is a submission made by a user in a subreddit. A post can, in turn, attract comments.}
\label{fig:reddit}
\end{figure}

Existing research has not adequately addressed the complexity of these discussion structures. Studies have either focused on specific domains like political discourse or toxic conversations~\cite{gonzalez2010structure,saveski2021structure}, or examined a limited set of features such as content type~\cite{yy}. While these studies provide valuable insights, they fail to capture the broader dynamics of online discussions across various contexts. To address these limitations, we propose a comprehensive framework that distinguishes between local features (post-specific attributes) and global features (community-level characteristics) of a discussion post.

Local features include post quality, medium (text, image, or video), sentiment, and length. These attributes can significantly influence how users initially engage with a post and shape the subsequent discussion. For instance, a high-quality, emotionally charged post might spark more intense debate, leading to a deeper discussion tree. Global features, on the other hand, include topics and communication norms within communities~\cite{aragon2017generative}. These community-level characteristics can set the tone for discussions and influence user behavior across multiple posts. For example, a subreddit focused on scientific topics might foster more in-depth, evidence-based discussions compared to one centered on casual conversation. We argue that analyzing both global and local features is essential for a comprehensive understanding of online discussions. This dual perspective offers a more nuanced understanding of the dynamics shaping online conversations, considering both the immediate context of a post and the broader community environment in which it exists.

Thus, we quantify the impact of both global and local features of Reddit posts on discussion tree structures. Unlike previous studies that focus on building generative models to predict discussion tree growth~\cite{aragon2017generative}, we adopt a discriminative modeling approach instead. We use regression analyses to examine how various features contribute to the structural properties of discussion trees. This approach allows us to quantify the relative importance of different features and their interactions in shaping discussion structures, providing more interpretable and actionable insights.

We conduct our study using Reddit, a large social news aggregation and discussion website with tree-like discussion structures. Reddit's diverse ecosystem of communities (subreddits) provides an ideal platform for examining how different features influence discussions across various contexts. Figure 1 illustrates a Reddit post and its discussion tree. We consider three key structural elements: width, depth, and size, which collectively capture the overall shape and scale of the discussion.

Our research addresses the following questions:

\begin{itemize}
\item [{\bf RQ1:}]   What is the relative contribution of a discussion post's global and local properties in predicting the structural properties of the discussion tree? How do these patterns vary with the size of a discussion thread?

\item [{\bf RQ2:}]  How do specific properties of a discussion post in both global and local features\textemdash e.g., its topic, sentiment, novelty, content type, or quality\textemdash moderate the breadth, depth, and size of conversations?

\end{itemize}

% they have yet toinvestigated the comparative significance of post-related versus community-related factors in shaping discussion structures.

% Understanding the relative importance of post-related and community-related features is not only crucial for filling the knowledge gap in academic and practical contexts but also has practical implications. For instance, advertisers seeking to optimize their strategies within a limited timeframe must make decisions about whether to invest more time in revising posts or finding the right community for delivering their content. In such scenarios, it becomes imperative to determine which aspect—post or community—should be prioritized to achieve the goal of obtaining diverse feedback on a product.

% To answer our research questions, we conduct quantitative statistical analyses using \texttt{Reddit.com} data.
% e approach the research question quantitatively via various statistical analyses at varying granularities to explore the relative importance of global and local features and to pinpoint the effect-size of particular properties of the discussion post on the structure of the conversation tree while controlling for a variety of confounds.
By addressing our research questions, we aim to provide a comprehensive understanding of the factors shaping online discussions. Our findings have implications for both theoretical understanding of online discourse dynamics and practical applications in platform design and content moderation strategies.

Our results reveal that both local and global features contribute significantly to explaining structural variation in discussion trees. Local features, such as post content and sentiment, collectively explain more variation in the discussion trees' structural properties than global features. However, we observe significant heterogeneity in the impact of various features. Several global features, including the subreddit's topic, age, popularity, and content redundancy, also play crucial roles in understanding specific properties of discussion trees. Notably, posts with negative sentiment and those that are primarily text-based tend to generate deeper and wider discussion trees. Similarly, posts from subreddits focusing on politics, sports, local events, and business are more likely to spark extensive discussions.

These findings offer valuable insights into how community-wide (global) and individual (local) features impact online discussions. At the community level, our results illuminate overarching patterns that contribute to the formation of engaging and meaningful discussions. Platform designers and community moderators can leverage these insights to foster constructive discourse on a larger scale, potentially by adjusting recommendation algorithms or community guidelines. On an individual level, our study provides users with actionable information for tailoring their posts to create the type of discussion they intend to have. For instance, users aiming for in-depth discussions might focus on crafting text-heavy posts on specific topics known to generate deeper conversations.

By bridging the gap between theoretical understanding and practical application, our research contributes to the ongoing efforts to improve the quality and effectiveness of online discussions. These insights can guide the development of more sophisticated content moderation tools, help refine user interfaces to encourage productive conversations, and empower users to engage more effectively in online communities.

\section{Related Work}

Our work builds upon a rich body of literature on studying online discussion threads or trees \cite{medvedev2019anatomy}, particularly focusing on generative models developed to explain their structural properties (see a comprehensive survey by \citet{aragon2017generative}). These models typically employ factors such as popularity, novelty, and reciprocity to analyze and predict the size, depth, and degrees of discussion trees \cite{kumar2010dynamics,wang2012user,
gomez2013likelihood,aragon2017thread}. Some studies also incorporate a root-bias feature, which explicitly differentiates the root node (i.e., the original poster) and other commentators \cite{gomez2008statistical,lumbreras2016automatic,aragon2017thread}. \citet{backstrom2013characterizing} explored how the arrival patterns of users influence the size of the discussion trees. \citet{lumbreras2016automatic} investigated the distinct roles users take and their effects on the shape and size of the discussion trees. \cite{kim2023predicting} examined the continuity of existing Reddit discussion trees by predicting whether a new comment will arrive or not. \citet{nishi2016reply} analyzed reply trees on Twitter with a branching process model, particularly on the effects of segments without branching. \citet{medvedev2019modelling} predicted the growth of discussion trees on Reddit over time using the Hawkes process. \citet{horawalavithana2022online} used a generative algorithm to forecast the discussion tree structure of groups of Reddit posts that share similar topics. Another line of research aimed to explain the depth of the tree, which is the longest path from the root of the tree to a leaf, or the size of the tree, which is the total number of participating comments \cite{kumar2010dynamics,backstrom2013characterizing}. Tree depth and size are proxies for general interest in the post since longer and larger trees indicate more participants. 
% \citet{backstrom2013characterizing} also examined factors explaining user reentry--when a user comments again on a post they previously interacted with. \cite{zhang2018characterizing} models the growth trajectory of a discussion post as well as how likely a post will result in an antisocial event (e.g., blocking). 
However, to the best of our knowledge, there is no prior research on the interplay between global versus local features of discussion trees on their size, depth, and width.

Most closely related to our work is a paper by \citet{zhang2018characterizing} that examines a  range of different Facebook \textit{Pages}, which are sub-communities that are in many ways similar to subreddits. Using a series of network features, they predict the discussion tree trajectory as well as how likely a post will result in an antisocial event (e.g., blocking) \cite{zhang2018characterizing}. While this work and others provide inspiration for using network features to quantify conversation, they do not consider how communities, such as different Facebook Pages, might contribute to the discussion tree or conversation network\cite{yu2023large,gamba2024exit}.

While local properties of posts have been extensively analyzed in previous literature, how discussion tree structures differ by their global features of each subreddit, to the best of our knowledge, have not been considered. Our work bridges this gap by analyzing the macro-level structures of discussion forums at the subreddit level. 
We theorize that subreddits may embody distinct characteristics in the discussion tree structure, perhaps due to the theme of the subreddit. Besides obvious topical differences (e.g., r/leagueoflegends, r/harrypotter), the intent of each subreddit may also be different. Some operate as Quora Q\&A platforms (e.g., r/AskReddit, r/ELI5 ``Explain it Like I'm 5'') and others operate as discussion forums soliciting advice or questions (e.g., r/IAmA, r/relationship\_advice). Many forums are also dedicated to adult, not-safe-for-work content. Our research bridges the gap in studying conversational threads by focusing on how global and local properties together explain conversational tree structural properties.

Our motivation for this work is to draw a parallel between the hierarchical relationship of a subreddit with the larger platform of Reddit and the relationship of individual trees within a forest ecosystem. Just as a forest houses a multitude of distinct trees, Reddit hosts numerous specialized communities known as subreddits. Each subreddit focuses on a particular subject, theme, or interest, acting as a hub where like-minded individuals congregate, share ideas, and engage in discussions. Collectively, subreddits play a crucial role in shaping the user experience and the overall culture of Reddit. As such, in this work, we emphasize the distinction between \textit{local} features--features specifically pertaining to each individual post--and \textit{global} features--features describing the subreddit as a whole. The global and local features together offer a clearer explanation of discussion tree structures. We metaphorically compare posts or conversations in a subreddit with branches on a tree, which therefore makes Reddit a \textit{forest of trees}.

The dual focus on global and local features is supported by other works on characterizing online communication. \citet{zhang2017community} characterized user engagement patterns of different subreddits based on how their collective identity is shaped by the distinctiveness and temporal dynamics of the subreddit. \citet{lambert2022conversational} studied how characteristics of the subreddit and the users, amongst other factors, can explain whether a Reddit conversation will continue after an adverse event, defined as when a moderator removes a community-violating comment. They found that both global and local features impact the resilience of conversation continuing after such adverse events. For instance, conversations in more populated subreddits (a global feature) are more resilient and prosocial following an adverse comment. Additionally, users with a history of removed or toxic content (a local feature) would likely end up halting the thread. \citet{coletto2017automatic}, in their work on detecting controversy in communities, found that local network patterns (motifs) can sufficiently predict controversy. In another work, \citet{shugars2019keep} demonstrated the importance of local features, showing that tweet-specific features, especially content with very negative emotions, can spark sustained user engagement in terms of follow-up discussions. Another study by \citet{yy} linked the size, virality, and responsiveness of subreddit posts to local features of the post's textual properties and user activity patterns. They also suggest that global properties may play a role, as news-related and image-based subreddits are more likely to have large and responsive conversations, while discussion-based subreddits tend to have more viral conversations.

In the broader field of computer-supported cooperative work (CSCW), related studies have examined other aspects of conversational structure, such as the impact of comment display formats on user retention~\cite{budak2017threading} and the role of social catalysts in online interactions~\cite{saveski2021social}. By synthesizing these diverse strands of research and introducing a novel framework that considers both global and local features, our work aims to provide a more comprehensive understanding of the factors shaping online discussion structures across different community contexts.

% other novelty
% - topic of subreddits
% - redundancy of subreddits
% - video
% - nsfw (adult content)
% - other simple text characteristics

\section{Data}
To answer our research questions, we use data from \texttt{Reddit.com}, a large social news aggregation, web content rating, and discussion website. The scale and popularity of Reddit as a discussion platform make it a highly representative testbed to explore and quantify the connection between the structure of online discussions and the properties of the individual posts themselves. 
% We perform various statistical analyses at varying granularities, including exploratory analysis, Partial R-square analysis, and regression analysis, to pinpoint the effect size of particular properties of the discussion post on the structure of the conversation tree while controlling for a variety of confounds.

\subsection{Data Description}
We sourced our dataset and the list of subreddits using the  PushShift API.
% \\footnote{\url{https://pushshift.io}. We are aware of the potential issues with data sourced from PushShift~\cite{gaffney2018caveat}. However, we used it in this paper due to the prevalence of its use in previous work.}. 
Our data consists of posts from a set of 500 subreddits that were randomly selected from 5,871 public, English language, safe-for-work subreddits listed on PushShift that had at least 1,000 discussion posts.\footnote{There were a total of 1,067,471 subreddits based on the PushShift data dump from 5/2018 (cf. \url{https://files.pushshift.io/reddit/subreddits/older_data/}). For our analyses, we consider subreddits created before 1/1/2018 (to further control for age) and that had at least 1000 discussions with at least one comment in 2018 (in other words, they were not inactive subreddits). Out of this subset, we further restricted our analyses to English language and safe-for-work subreddits. This led to the final set of 5,871 subreddits that we analyzed.} We divided these subreddits into ten strata (deciles of the distribution) based on the number of subscribers and then randomly sampled 50 subreddits from each stratum.

Next, we randomly sampled a full year's worth of data from 2018 (an uninterrupted year pre-COVID) from these 500 subreddits.\footnote{We will add the full list of the 500 subreddits into the Appendix upon acceptance of the paper.} We sampled from this fixed time window to control for time effects, as some subreddits are significantly older than others. Then, from each subreddit, we randomly sampled 500 posts (discussion trees). We ensured that these randomly sampled discussion trees were not empty and that each had comments. So, our final dataset consists of 250,000 ($500\times 500$) posts with a total of 3,574,020 comments. Unfortunately, several comments and posts were missing due to deletion or data collection issues, so the discussion trees generated from the comments were disconnected. Hence, we focused our analyses on the largest connected component of the discussion tree. Next, we labeled the 500 subreddits with topics based on the advertiser categories available as part of the PushShift Reddit dataset via manual annotation by two research assistants. The final list of the twelve subreddit topics and their summary statistics are shown in Table~\ref{tab:Subreddit_desc}.

\begin{table}
\caption{Subreddit topics and summary statistics. {\it Note:} We perform our analyses at the level of posts, so we list the total number of posts that were categorized into different topics.}

    \begin{tabular}{c|r|l}%p{4.5cm}}
        \toprule
        \textbf{Topics} & \textbf{\# Posts}  & \textbf{Sample Subreddits} \\
        \midrule
         Games & 56,500 & r/DotA2. r/gachagaming, r/DarkSouls2 \\
        \hline
        Entertainment & 55,500 & r/WeAreTheMusicMakers, r/gameofthrones, r/NolanBatmanMemes  \\
        \hline
        Lifestyle   & 53,500 &  r/KitchenConfidential, r/Bonsai, r/SkincareAddiction   \\
        \hline
        Technology & 25,500& r/GooglePixel, r/iOSthemes, r/AskProgramming  \\
        \hline
        Local & 14,500&  r/greece, r/greenville, r/wisconsin \\
        \hline
        Sports  & 11,500 & r/angelsbaseball,  r/ussoccer, r/Swimming \\
        \hline
         Health  & 8,000 &  r/ptsd, r/schizophrenia, r/bodyweightfitness \\
        \hline
         Business  & 7,500 & r/personalfinance, r/DisneyPinSwap, r/salesforce\\
        \hline
         Automotive  & 6,500 & r/Cartalk, r/Porsche, r/Miata\\
        \hline
        College  & 5,500  & r/UofT, r/uwaterloo, r/nyu \\
        \hline
        Politics  & 4,000  & r/policydebate, r/Socialism\_101, r/progun\\
        \hline
        Miscellaneous  & 1,500  & r/FreeKarma4You,  r/whatisthisthing,  r/darknet \\
        \bottomrule
    \end{tabular}
  \label{tab:Subreddit_desc}
\end{table}

\subsection{Feature Generation}
We generate various global and local features of the discussion posts. As described earlier, the \textit{global} features of a post are the features encoding the broader context in which a post was made, whereas the \textit{local} features capture the post's attributes. Guided by several previous studies in our choice of features of discussion posts, we use a broad set of features, including the topic of conversation~\cite{gonzalez2010structure,romero2011differences}, the age of the online community, the type of content~\cite{cha2012spread}, emotional valence~\cite{berger2012makes,stieglitz2013emotions}, and several textual features of the post.

\subsubsection{Global features:}
We consider several key global features of the discussion posts, which describe the properties of the community or subreddit in which a given post was made.

\begin{figure} %[htbp]
\centering
% Note: text {\it Width} is the {\it Width} of both columns
\includegraphics[width=0.9\textwidth]{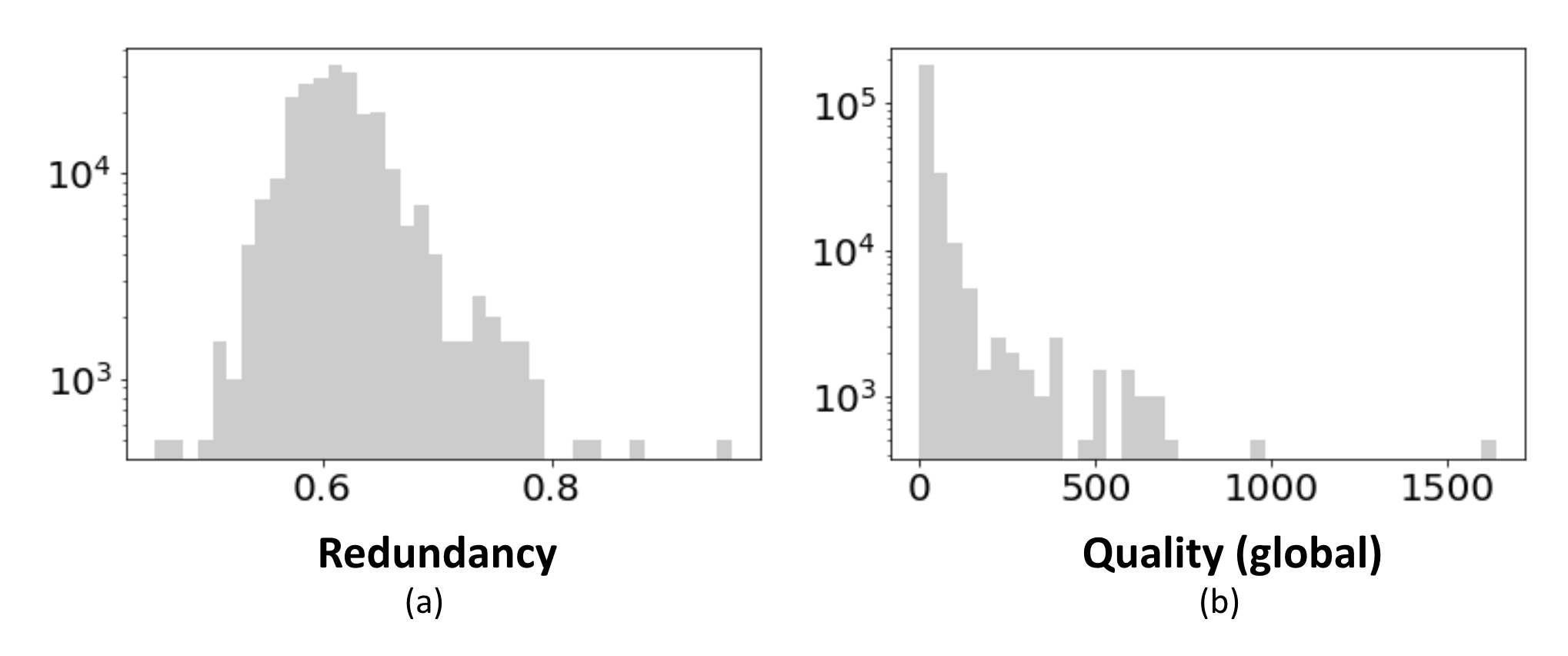}
\caption{Empirical distribution of two {\it global} features: ``average {\bf quality} of posts'' ($\mu$=59.6, $\sigma$=134.1, min=1.1, max=1634.7, median= 21.1) and the ``{\bf redundancy} of content'' ($\mu$=.62, $\sigma$=.05, min=.45, max=.96, median= .61) .}
\label{fig:summaryGlobal}
\end{figure}

\begin{itemize}
    \item {\bf Topic:} This feature encodes the broad category the subreddit is classified under. We retrieved the ten subreddit advertiser categories sourced from PushShift and added two more categories, politics and miscellaneous, for completeness. Unfortunately, many of these categories were missing from the data. Hence, we asked two research assistants to annotate the data and verify the labels to classify each subreddit as one of the twelve topics. The manual annotation process involved reading the description and content of the subreddit and then discussing it until an agreement was reached between the annotators. Note that a given subreddit can only be part of one category. See Table~\ref{tab:Subreddit_desc} for summary statistics and examples of each topic.
    
    \item {\bf Popular:} This is a binary feature that denotes the popularity of a subreddit. In our definition, a subreddit is deemed popular if it has more than the average number of subscribers among the subreddits we extracted.
    
    \item {\bf Novelty:} We consider how new a subreddit is since the anatomy of a post's discussion tree also depends on how old a given subreddit is. Older subreddits have had a chance to gather more subscribers since they have been around longer, and they also have more posts on average. We define a subreddit as new if it is at most a year old. Since our data is from 2018, a subreddit is considered new if it was created at the earliest in 2017.
    
    \item {\bf Average quality of posts:} Quantifying the quality of discussions in a subreddit is challenging and can be subjective. Hence, we operationalize the quality of conversations in a subreddit using \textit{karma}, a metric provided by the Reddit platform itself and one that has also been used in previous research~\cite{zayats2018conversation}. The \textit{karma} of the post is based on how many upvotes versus downvotes it received. We then define the quality of discussions in a subreddit as the average {\it karma} of all the subreddit posts. The empirical distribution of the average {\it karma} is shown in Figure~\ref{fig:summaryGlobal}.
    
    \item {\bf Redundancy of content:} A critical determiner of user participation in online community discussions is how redundant are the posts in a subreddit. To this end, we developed a metric to measure how similar each post is to the others in the same subreddit. We first calculate post embeddings as the average of their Word2Vec embeddings \cite{mikolov2013distributed}, then we aggregate subreddit embeddings as the TF-IDF weighted average of their post embeddings\cite{yu2023unique}. The redundancy score is computed as the mean cosine similarity between every post embedding and the subreddit embedding. A higher redundancy score reflects more repetitive content in the subreddit.
    % So, we develop a new metric to quantify a Subreddit's redundancy using word2vec, a vector space model of word semantics~\cite{mikolov2013distributed}. First, we calculate the mean semantic vector ($MSW_i$) for a given Subreddit $i$ by computing the term frequency-inverse document frequency (TFIDF) weighted average of the word2vec embeddings of all the posts in that Subreddit. Next, we calculate the semantic vector $SW_j$ for each post $j$ in that Subreddit. Finally, we compute the cosine similarity between the semantic vectors for each post in a Subreddit $SW_j$ and the mean semantic vector $MSW_i$, and we average those numbers. Essentially, the mean cosine similarity that we just computed quantifies how semantically similar a given post is to the average post in that Subreddit. The higher the similarity score, the more repetitive the content of discussions in that Subreddit and vice-versa. 
    The empirical distribution of the feature is shown in Figure~\ref{fig:summaryGlobal}. As can be seen, most subreddits are neither highly redundant nor highly varied in their content.
\end{itemize}

\begin{figure}
\centering
% Note: text {\it Width} is the {\it Width} of both columns
\includegraphics[width=0.9\textwidth]{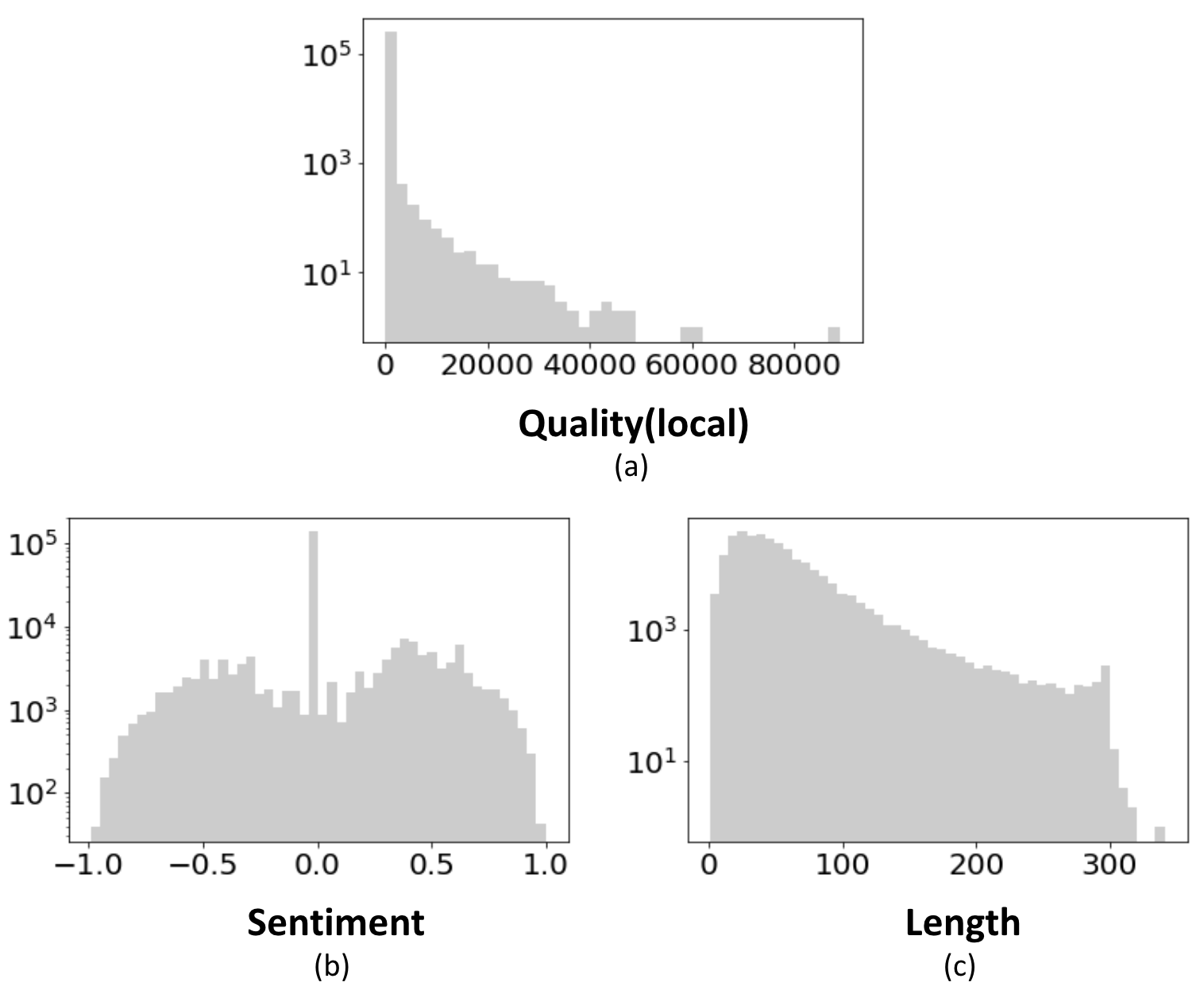}
\caption{Empirical distribution of the non-binary {\it local} features: ``{\bf quality}'' ($\mu$=68, $\sigma$=696.9, min=0, max=88908, median= 6), ``{\bf length}'' ($\mu$=50.5, $\sigma$=37.7, min=1, max=341, median= 41), and ``{\bf sentiment}'' ($\mu$=.055, $\sigma$=.322, min=-.987, max=.999, median= .000).}
\label{fig:summaryLocal}
\end{figure}

\subsubsection{Local features:}
Next, we generate various local features that describe a specific post.

\begin{itemize}
    \item {\bf Quality:} This is the direct analog of the  {global} quality feature, with the key difference that here it is just the {\it karma} of that particular post.
    \item {\bf Text:} Many posts on Reddit contain multimedia content, e.g., images, gifs, audio, and video. This binary feature indicates if the content of the post is text-only. We use this feature since users may interact differently with multimedia content compared to text-only content. For example, posts with texts may prompt more discussion instead of posts with only a meme image.
    
    \item {\bf Video:} This binary feature indicates whether the post contains video content. 
    \item {\bf Adult:} This binary feature indicates whether this post is marked not-safe-for-work (NSFW) and could contain adult content. Even though we excluded NSFW subreddits, there could still be posts in non-NSFW subreddits that are tagged as NSFW. For the same reason above, we expect users to interact to various modalities of posts differently.
    
    \item {\bf Sentiment:} We use the VADER algorithm~\cite{hutto2014vader} to compute the sentiment of the post on a scale of from $-1$ to $+1$ with $-1$ indicating the most negative sentiment and $+1$ being the most positive sentiment. 
    \item {\bf Contains numbers:} A binary feature indicating whether the post contains numbers, e.g., the feature would take a value 1 for the post ``Happy New Year 2020!''.
    \item {\bf Length:} This feature codes the number of characters in the post. One might expect longer posts to be more nuanced and thereby draw much interest from users. For example, The post ``Happy New Year!'' has a Length of 15 (including spaces).
    \item {\bf All uppercase:} This is also a binary feature that fires if the post is written in all uppercase letters. Typically, all uppercase has the connotation of shouting or screaming.\footnote{\url{https://en.wikipedia.org/wiki/All_caps}} So, we conjecture that the tone of the post could also determine the structure of the conversation that happens.
\end{itemize}
Figure~\ref{fig:summaryLocal} shows the summary statistics of the non-binary {\it local} features. As can be seen, the length of posts is usually under 300 characters, and most posts are of low quality. Further, most of the posts are neutral in terms of sentiment, and the average sentiment of posts in our dataset is slightly positive ($\mu$=.055).

\begin{figure}[htbp]
\centering
% Note: text {\it Width} is the {\it Width} of both columns
\includegraphics[width=.9\textwidth]{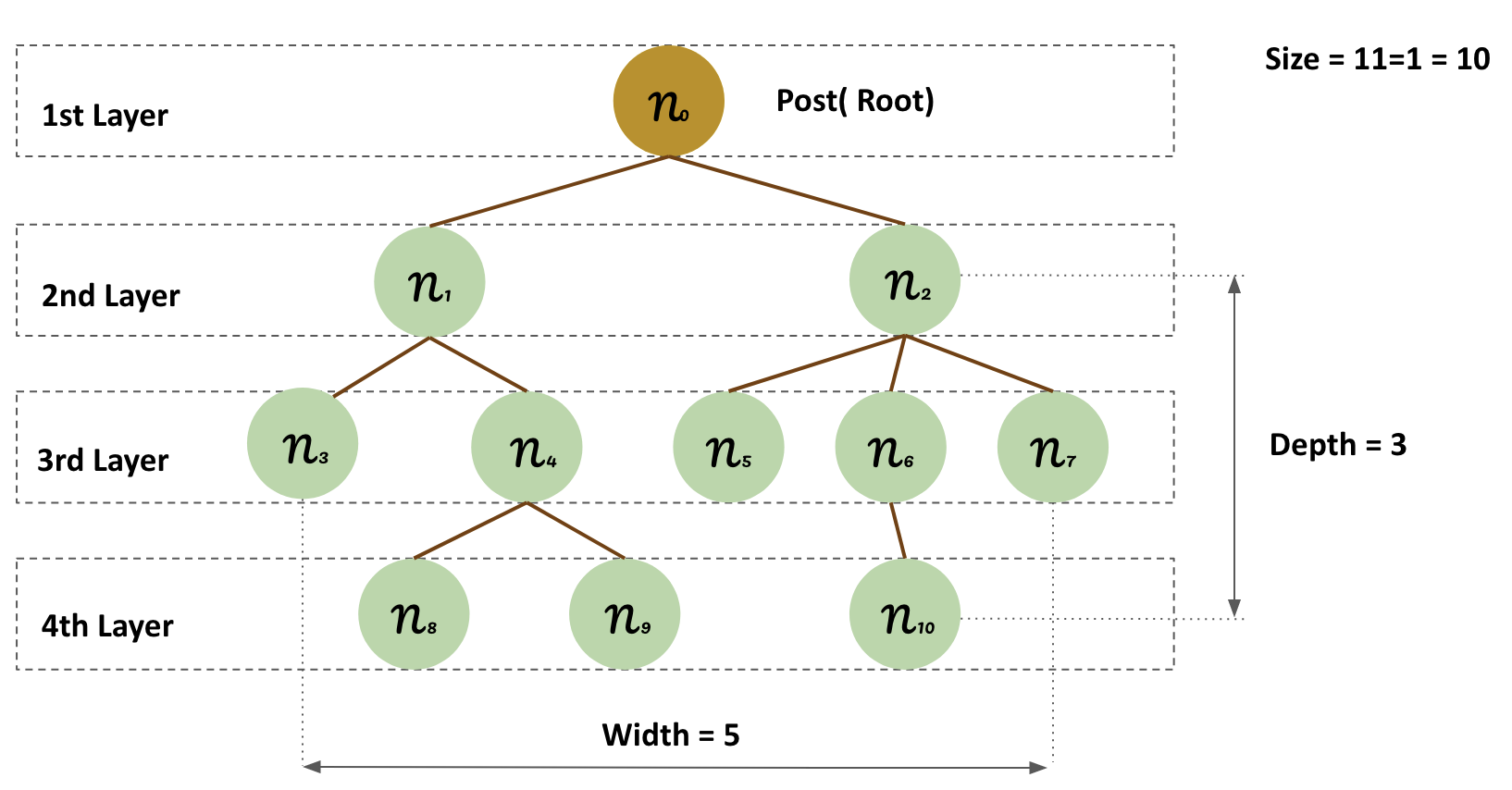}
\caption{Illustration of a discussion along with its various structural properties.}
\label{fig:dtree_prop}
\end{figure}

\subsection{Structural Properties of the Discussion Tree}
The comments made by users on a given post constitute a {\it threaded} structure and can be represented as an undirected graph $T =(V, E)$ with $V$ vertices and $E$ edges~\cite{aragon2017thread}. Since our data has missing comments (either deleted or missing otherwise), we use the comment tree's largest connected component for our analysis in this paper. We quantify these discussion trees' structures using several intuitive and commonly used properties of the underlying graph structure. The various structural properties are described in detail below and are shown graphically in Figure~\ref{fig:dtree_prop}.

% \begin{itemize}
% \item {\bf Size:} 
\paragraph{Size} First and foremost, the most important structural property of a discussion tree is its size. In our case, this is equal to the total number of comments made on that post. This structural property has also been previously used by~\cite{yy} to study user participation in discussions on Reddit. A tree's size is a very generic property as it captures the interest or buzz surrounding that post, but it says nothing about the conversations' actual structure. For example, consider two scenarios: one in which two users reply to each other's comments, leading to a very deep and narrow tree, and another in which several users respond to the original post, leading to a wide but shallow tree. Both these patterns of discussion can have the same number of comments leading to the exact size of the discussion tree, but they have entirely different conversation dynamics. Hence, as we describe next, we also consider other structural properties of a discussion tree, particularly its width and depth.

% \item {\bf Width:} 
\paragraph{Width}
The width of a discussion tree is simply the width of the broadest layer of the tree. It encodes the most expansive conversation in the entire discussion tree. It is also not limited to just the replies to the original post, as users can reply to a reply in nested discussion tree structures. Wider discussion trees typically have various sub-conversations happening simultaneously or contain a wide variety of opinions.

% \item {\bf Depth:} 
\paragraph{Depth}
The depth of a discussion tree is defined as the depth of the deepest branch of the tree. It can potentially allude to how controversial or contentious a given post and its follow-up comments are. Very niche conversations can also lead to very deep conversation trees.
% \end{itemize}

\begin{figure}

\centering
% Note: text {\it Width} is the {\it Width} of both columns
\includegraphics[width=0.9\textwidth]{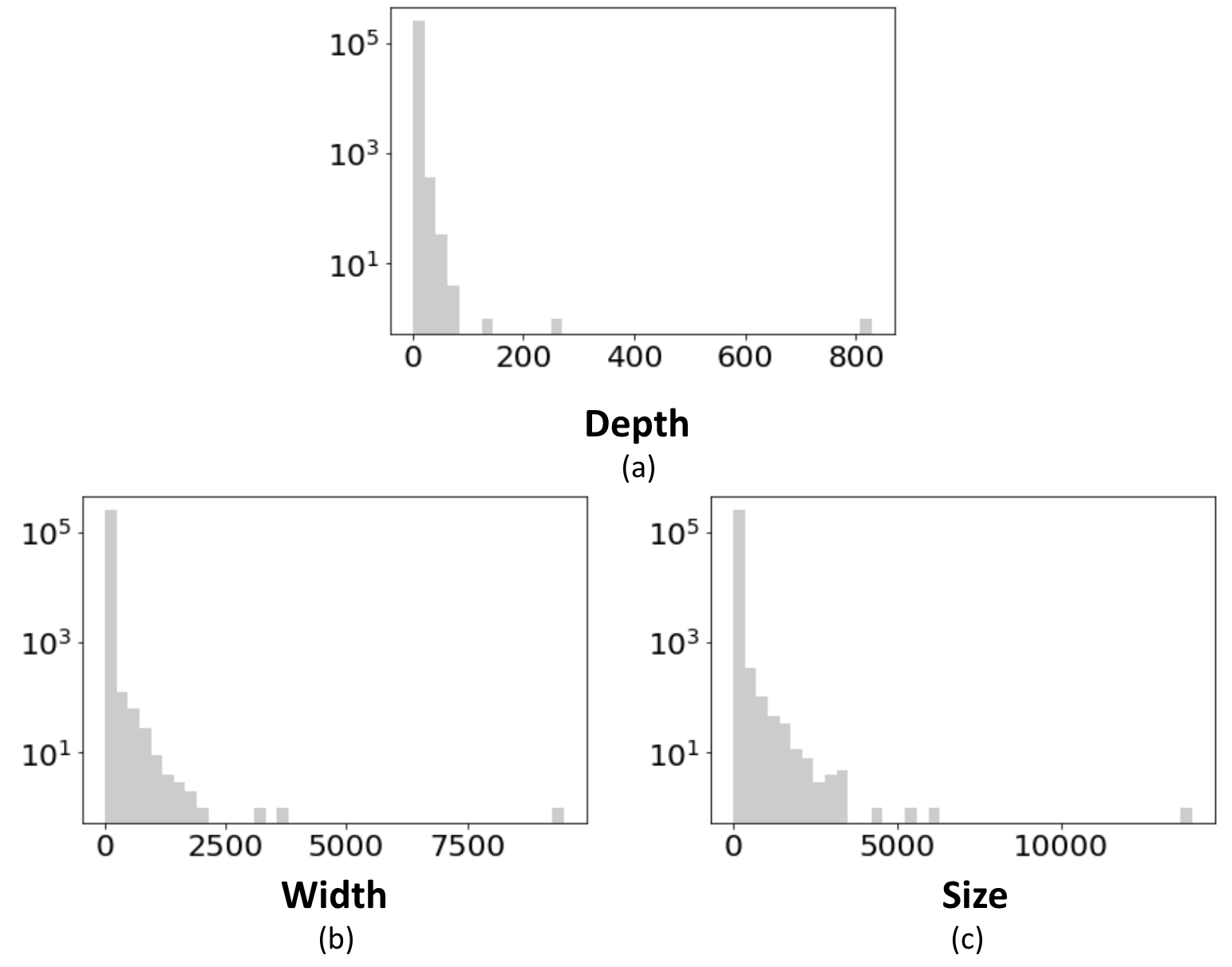}
\caption{Empirical distribution of the various structural properties of the discussion trees. ``{\bf Size}'' ($\mu$=14.3, $\sigma$=59.7, min=1, max=14001, median=6), ``{\bf Width}'' ($\mu$=6.24, $\sigma$=29.90, min=1, max=9499, median=3), ``{\bf Depth}'' ($\mu$=3.38, $\sigma$=3.32, min=1, max=828, median=3).}  
\label{fig:dTree}
\end{figure}

The empirical distribution of the various structural properties of the discussion trees is shown in Figure~\ref{fig:dTree}. As can be seen, the size and width variables are overdispersed, giving rise to a long-tailed distribution. However, the depth variable falls off a little sharply, leading to thinner tails of the distribution, thereby showing a lack of relatively long discussion chains in our dataset.

\section{Empirical Models \& Results}
Our empirical analyses proceed in two steps. First, we assess the relative importance of the global and local features collectively as a group. How much do these groups of features contribute to the variability in the discussion trees' structural properties? Are the breadth, depth, and volume of discussions driven primarily by the post's properties, or are they driven by the subreddit properties in which those conversations happen? Once we have quantified the impact of global and local features as a group, we proceed next to assessing individual features' importance in determining the structure of conversations on Reddit.

All our analyses center around estimating the empirical model:

\begin{equation}
% \nonumber
\texttt{StructuralProp}_k = \alpha + \sum_i \texttt{GlobalFeatures}_{ik} %&& \\
+ \sum_j\texttt{LocalFeatures}_{jk} + {\epsilon_k},
\label{eq:reg}
\end{equation}
The dependent variable $\texttt{StructuralProp}_k$  encodes the various structural properties (width, depth, and size) of the $k^{th}$ post. The independent variables $\texttt{GlobalFeatures}_{ik}$ and $\texttt{LocalFeatures}_{jk}$ encode the various {\it global} and {\it local} features for the $k^{th}$ post. The subscript $i$ indexes the list of all the global features, and similarly, the subscript $j$ indexes the set of local features that were described earlier. Finally, $\alpha$ represents the intercept term: the idiosyncratic variation in the structural tree properties that is not explained by the {\it global} or {\it local} features is absorbed by the Gaussian error term $\epsilon_k$.

Several of our variables are highly skewed (cf. Figures~\ref{fig:summaryGlobal}, \ref{fig:summaryLocal}, and \ref{fig:dTree}), which is problematic for regression models since they assume the dependent variable to be distributed according to a Gaussian Distribution \cite{angrist2008mostly}. One common solution for this problem, and one that we also adopt in this paper, is to log-transform the skewed variables $x$ as $\log(x+1)$.\footnote{To preserve the zero counts, we add one to the count before log-transforming it so that if $x=0$, then $log(x+1)$ is also 0.
% Log(0) is $-\infty$.
}

\subsection{Relative Importance of Global and Local Features}
To determine the relative importance of global and local features on the structure of discussion trees, we perform Partial R-squared analysis~\cite{kutner2005applied}, which compares the R-squared value\footnote{Also known as the coefficient of determination or amount of variance in the dependent variable that can be explained by the independent variable.} in the full model (cf. Equation~\ref{eq:reg}) with those of its two reduced models. One reduced model only contains the local features, and the other one only considers the global features. Comparing the full model with both of these reduced models allows us to differentiate the impact of global and local features.

The results, displayed in Table~\ref{tab:partial-r}, show that the local features explain significantly more variation in the structural properties of the discussion tree than the global attributes. For example, the local features explain 30.9\% of the variability in the discussion trees' width, while the global features only explain 2\% of the remaining variation. Similar trends also hold for the depth and size variables: the local features carry substantially more predictive power than the global features. These results, though interesting, are not entirely surprising. The local properties of the post influence the decision process of the users interacting with a post, thereby impacting the structure of follow-up discussions. That said, the global properties of a post also significantly impact the discussion structure, but their magnitude is between 6-16 times\footnote{Computed by taking the ratio of R-square of local and global features from Table~\ref{tab:partial-r}, e.g., $0.157/0.023=6.82$.} lower than the impact of local features. We also performed an adjusted partial R-square analysis as a robustness check, and our findings align qualitatively with the primary results.

%Paramveer: The statement below isn't correct. We use partial R2 to explain the relative variance explained by local vs global features. It is not because our dataset size is big---people use partial R2 on small datasets all the time.
%We opted to use the partial R-square metric because of the substantial size of our dataset. 
 
\begin{table}%[htbp]
\caption{Partial R-squared analysis results. The numbers represent the amount of variation explained using just the global and local features in predicting the width, depth, and size of the discussion trees.}
\begin{tabular}{l|lllll}
\toprule
\textbf{Structural Properties} $\rightarrow$& \textbf{Depth} & \textbf{Width} & \textbf{Size} \\
\midrule
% \hline
Local features & 0.157 & 0.309 & 0.307 \\
Global features  & 0.023 & 0.020 & 0.025\\
\bottomrule
\end{tabular}

\label{tab:partial-r}
\end{table}

\begin{figure}%[htbp]
\centering
% Note: text {\it Width} is the {\it Width} of both columns
\includegraphics[width=0.90\textwidth]{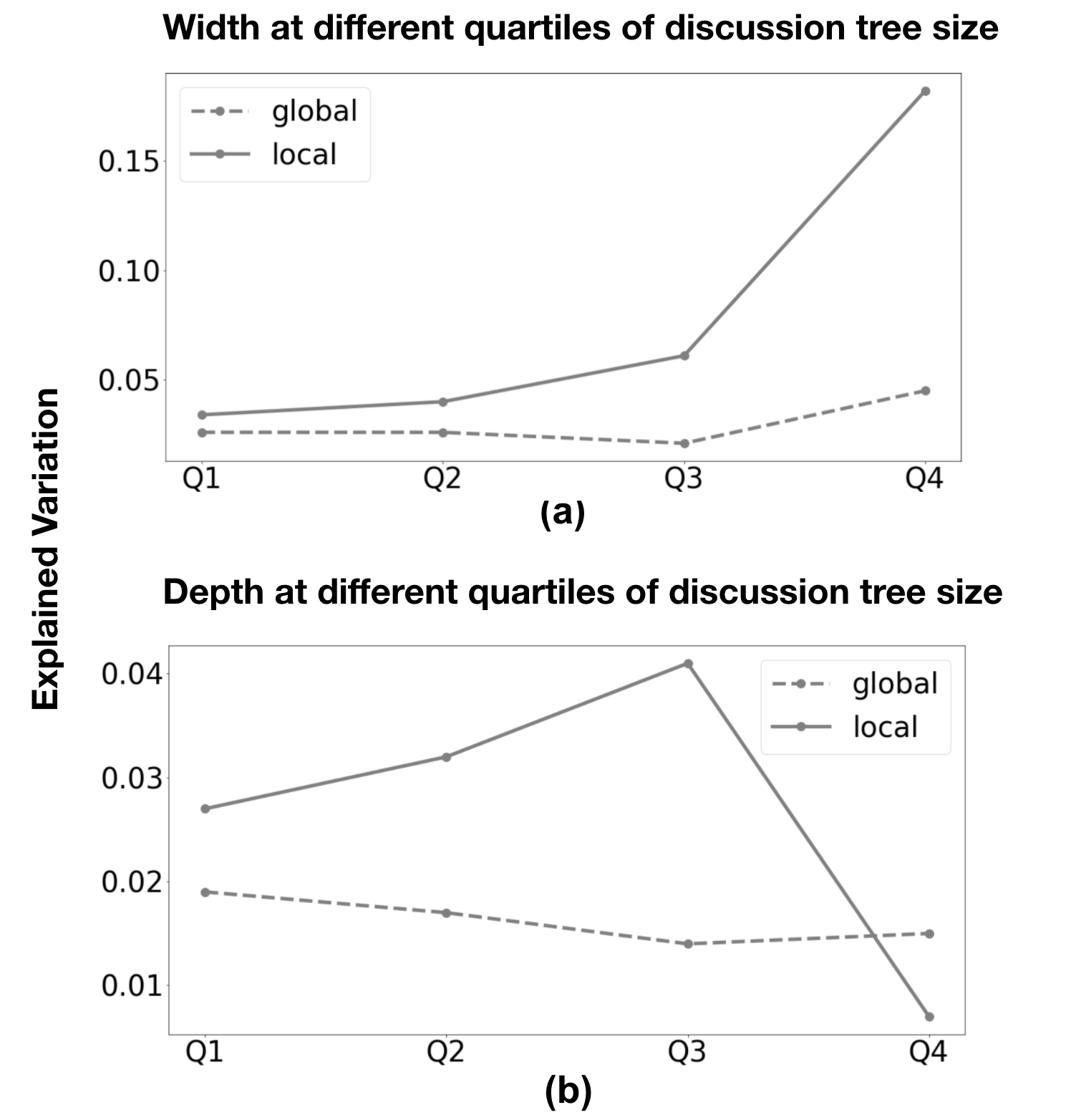}
\caption{Variation in discussion tree width (a) and depth (b) at different quartiles of discussion tree size, explained by global and local features. Q1: 0\%-25\%, Q2: 26\%-50\%, Q3: 51\%-75\%, Q4: 76\%-100\%.}
\label{fig:localGlobalHetero}
\end{figure}

To ensure the robustness of our findings, we conduct several additional analyses shown in Figure \ref{fig:localGlobalHetero}. We first perform an adjusted partial R-square analysis as a robustness check to reaffirm that local features exhibit a greater predictive capability than global features. Additionally, we replicate the partial R-squared analysis on different cross-sections of the data by quartiles of discussion tree sizes. Essentially, we subsample the discussion trees with at least ten comments in our dataset and stratified them into four groups (quartiles)-- 0\%-25\%, 26\%-50\%, 51\%-75\%, and 76\%-100\%-- based on their size. We find that across all the quartiles, the local features explain more variation in the discussion tree's width compared to the global features. Similarly, the local features explain more variation in the depth of the discussion trees, except for larger discussion trees (76\%-100\%), where the pattern is reversed. This reversal in pattern can be explained by the fact that as a discussion tree increases in size, the conversation's scope also becomes broader than what can be explained narrowly by the post's local properties alone. 
 
\subsection{Quantifying the Impact of Individual Features}
Next, we quantify the impact of specific important global and local features on the properties of the discussion trees. We first perform an exploratory study to visualize how particular features separate different discussion tree structures. To do so, we create a two-dimensional grid that categorizes the discussion trees according to their depth and width. Depth and width are two of the more critical structural properties since they add more nuance to the information regarding the sheer volume of discussions captured by the {\it size} variable. For instance, discussion trees with more depth could indicate a more focused to-and-fro conversation between the participants. Deeper conversation trees are also more likely to comprise stimulating discussions, making the users engage with the post by {\it replying} to it. In contrast, wider discussion trees indicate several sub-conversations happening simultaneously, potentially leading to a broader array of opinions being expressed or remarks being made.

\begin{figure*}[t]
\centering
% Note: text {\it Width} is the {\it Width} of both columns
\includegraphics[width=.90\textwidth]{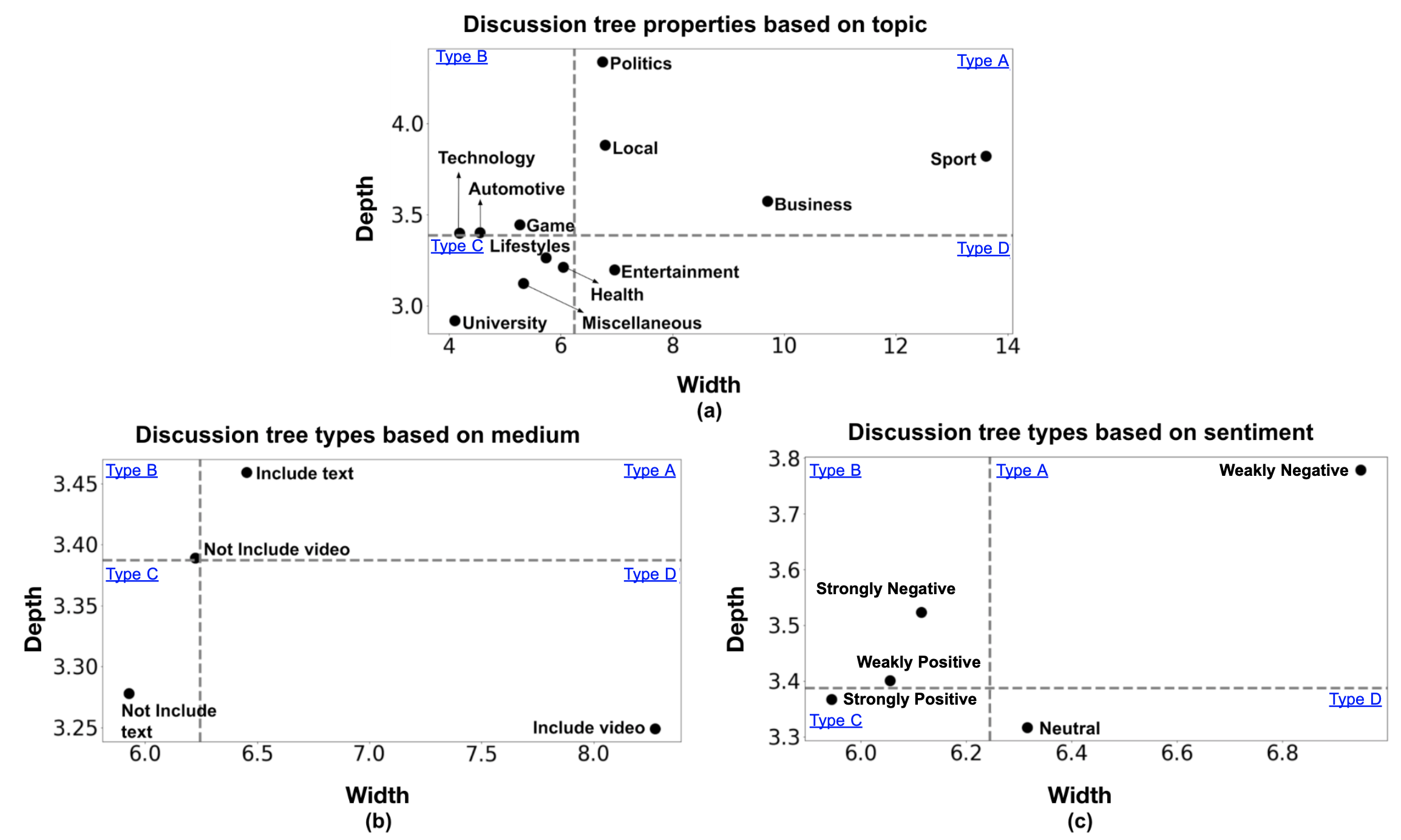}
\caption{Projection of various features onto the two-dimensional grid of width and depth of discussion trees. {\it Note:} 1) The dotted lines denote the mean of the width and depth of all the discussion trees; 2) Types A-D denote the discussion trees with higher than average width and depth, only depth, none out of width and depth, and only width, respectively; 3) Figure panel (a) plots the topics, panel (b) plots the multimedia text/video nature of the posts (c) plots the sentiment of the posts, and panel; 4) The ``sentiment'' in figure panel (c) is classified as either strongly positive $[0.5, 1]$, weakly positive $[0, 0.5)$, weakly negative $[-0.5, 0)$, or strongly negative $[-1, -0.5)$ according to their raw VADER \cite{hutto2014vader} sentiment scores from $-1$ to $+1$.}
\label{fig:grid}
\end{figure*}

Figure~\ref{fig:grid} illustrates the subreddits annotated according to some of their important post properties (both local and global), projected onto 2D space that characterizes the depth and width of their average discussion trees. To isolate interesting patterns, we further divide this grid into regions with higher or lower than average width and depth. This leads to four disjoint regions, shown by the dotted lines. Compared to the average width and depth, discussion trees in the Type A regions are both wider and deeper, Type B ones are less wide but deeper, Type C regions are both less wide and less deep, and Type D regions are wider but less deep.

Building on these exploratory results, we next perform a more rigorous regression analysis to assess the impact of particular post properties on the structure of discussions. Unlike the previous exploratory analysis, the regression analysis allows us to control for various features while quantifying a specific feature's marginal impact on the discussion trees' structure. We also control for the size of the discussion tree when assessing the impact of other features on the depth and width of the discussion tree ($\text{X}_{k}$ is the vector of control variables). To enhance the interpretability of the regression table, we also include standardized continuous variables in the regression. To further assess the robustness of our findings, we conducted a separate analysis using Reddit posts containing text. Our final results remained qualitatively consistent.

\begin{eqnarray}
% \nonumber
\texttt{StructuralProp}_k = \alpha + \sum_i \texttt{GlobalFeatures}_{ik} %&& \\
+ \sum_j\texttt{LocalFeatures}_{jk} + \sum\texttt{X}_{k}+{\epsilon_k},
\label{eq:reg11}
\end{eqnarray}

For the sake of simplicity and ease of interpretation~\cite{angrist2008mostly}, we choose the ordinary least squares (OLS) regression to estimate the specification in Equation~\ref{eq:reg11}. Although OLS regression assumes a linear relationship between the various local and global properties of the discussion tree and their structure, we acknowledge the possibility of non-linearities existing in this relationship and add a non-linear control variable, the size of the discussion tree.  Regarding the main variable (global and local features), our primary focus lies on examining the main effects rather than the higher-order effects typically captured by non-linear models. 
% Nonetheless, we also ran
% Of course, by using an OLS regression, we do not rule out the presence of non-linearities in the relationship between the various local and global properties of the discussion tree and the structure of those trees. We are, however, less interested in higher-order effects, which are typically captured by non-linear models. That said, we did run 
%non-linear Poisson and Negative Binomial regression models to explore alternative modeling strategies.  However, the results from these non-linear models did not yield significantly different directions and magnitudes of effect sizes when compared to the OLS regression model. 

We present the detailed estimation results from the OLS regression in Table~\ref{reg:regresstable}. To aid readability, we also provide a more concise summary of all our findings regarding the impact of various local and global features on the structure of discussion trees in Table \ref{tab:summary}. Below, we discuss several interesting observations from both the exploratory and the regression analyses regarding the impact of individual features on the structure of discussions. 

\begin{table}[htbp]
\caption{Regression estimates obtained by modeling Equation~\ref{eq:reg11}. {\it Note:} 1) Heteroskedasticity Robust standard errors are shown in parenthesis next to the coefficient, 2) ***$p < 0.01$, **$p < 0.05$, and *$p < 0.1$; 3) Intercept is removed for simplifying the interpretation; 4) `$l$' refers to local features and `$g$' refers to global features; 5) Green color refers to features that have a significantly positive effect on the dependent variables, while red color indicates a negative relationship. }
\label{reg:regresstable}
\begin{center}
\begin{tabular}{l|lll}
    \hline
    \hline
D.V.--\textgreater{} & \it{Depth}                       & \it{Width}                       & \it{Size}                    \\
    \hline
$\text{Quality}_l$                               & \textcolor{red}{-0.049***(0.001)}             & \textcolor{darkgreen}{0.051***(0.001) }             & \textcolor{darkgreen}{0.582***(0.002) }              \\
$\text{Text}_l$                                     & \textcolor{red}{-0.03***(0.001)}              & \textcolor{darkgreen}{0.046***(0.001)}              &\textcolor{darkgreen}{ 0.53***(0.004) }               \\
 $\text{Video}_l$            & \textcolor{red}{-0.04***(0.005)}              & \textcolor{darkgreen}{0.052***(0.005)}              & \textcolor{darkgreen}{0.112***(0.016) }              \\
 $\text{Adult}_l$            & \textcolor{darkgreen}{0.03***(0.007) }              & \textcolor{red}{-0.012*(0.007)}               & \textcolor{darkgreen}{0.177***(0.018) }              \\
 $\text{Sentiment}_l$        &\textcolor{red}{ -0.006***(0.001) }            & \textcolor{darkgreen}{0.004***(0.001) }             & \textcolor{red}{-0.024***(0.002) }             \\
 $\text{Numbers}_l$          & 0.001(0.001)                 &\textcolor{red}{ -0.003**(0.001)}              & \textcolor{darkgreen}{0.04***(0.004)}                \\
 $\text{Length}_l$           & \textcolor{darkgreen}{0.004***(0.001) }             & \textcolor{red}{-0.003***(0.001) }            & \textcolor{darkgreen}{0.073***(0.002) }              \\
 $\text{Upper}_l$            & \textcolor{red}{-0.016***(0.006) }            & \textcolor{darkgreen}{0.02***(0.006) }              & \textcolor{darkgreen}{0.018(0.017)}                  \\
  $\text{Redundancy}_g$     & \textcolor{red}{-0.005***(0.001)}             & \textcolor{darkgreen}{0.004***(0.001) }             & \textcolor{red}{-0.029***(0.002) }             \\
  $\text{Quality}_g$        & \textcolor{red}{-0.009***(0.001) }            & \textcolor{darkgreen}{0.012***(0.001) }             & \textcolor{red}{-0.028***(0.002) }             \\
  $\text{Novelty}_g$                                & \textcolor{darkgreen}{0.014***(0.002)}              & \textcolor{red}{-0.016***(0.002)}             & \textcolor{red}{-0.12***(0.007)}               \\
  $\text{Popular}_g$                                &\textcolor{red}{ -0.033***(0.002)}             & \textcolor{darkgreen}{0.041***(0.002) }             & \textcolor{darkgreen}{0.083***(0.006)  }             \\
   $\text{Sports}_g$         & \textcolor{darkgreen}{0.273***(0.004) }             & \textcolor{darkgreen}{0.435***(0.004) }             & \textcolor{darkgreen}{1.92***(0.009) }               \\
   $\text{Lifestyle}_g$      & \textcolor{darkgreen}{0.302***(0.003) }             & \textcolor{darkgreen}{0.418***(0.003) }             & \textcolor{darkgreen}{1.638***(0.005) }              \\
   $\text{Entertainment}_g$  &  \textcolor{darkgreen}{0.289***(0.003)  }            &\textcolor{darkgreen}{ 0.427***(0.003) }             & \textcolor{darkgreen}{1.6***(0.005) }                \\
  $\text{University}_g$     & \textcolor{darkgreen}{0.303***(0.005) }             & \textcolor{darkgreen}{0.434***(0.005) }             & \textcolor{darkgreen}{ 1.484***(0.011) }              \\
   $\text{Games}_g$          & \textcolor{darkgreen}{0.308***(0.003) }             & \textcolor{darkgreen}{0.4***(0.003) }               & \textcolor{darkgreen}{1.737***(0.005) }              \\
   $\text{Business}_g$       & \textcolor{darkgreen}{0.302***(0.004) }             & \textcolor{darkgreen}{0.4***(0.004) }               & \textcolor{darkgreen}{1.781***(0.012) }              \\
   $\text{Local}_g$          & \textcolor{darkgreen}{ 0.28***(0.004)}               & \textcolor{darkgreen}{0.452***(0.003) }             & \textcolor{darkgreen}{1.928***(0.008) }              \\
   $\text{Technology}_g$     & \textcolor{darkgreen}{0.346***(0.003)  }            & \textcolor{darkgreen}{0.363***(0.003) }             & \textcolor{darkgreen}{1.696***(0.006)  }             \\
   $\text{Health}_g$         & \textcolor{darkgreen}{0.262***(0.004)  }            & \textcolor{darkgreen}{0.464***(0.004) }             & \textcolor{darkgreen}{1.616***(0.01) }               \\
   $\text{Automotive}_g$     & \textcolor{darkgreen}{0.337***(0.004) }             & \textcolor{darkgreen}{0.371***(0.004)  }            &\textcolor{darkgreen}{ 1.687***(0.009) }              \\
   $\text{Politics}_g$       &\textcolor{darkgreen}{ 0.248***(0.009)  }            & \textcolor{darkgreen}{0.515***(0.008)}              & \textcolor{darkgreen}{1.623***(0.021) }              \\
   $\text{Miscellaneous}_g$  & \textcolor{darkgreen}{0.344***(0.005)  }            & \textcolor{darkgreen}{0.382***(0.005) }             &\textcolor{darkgreen}{ 1.795***(0.015) }              \\
Size\_log                             & \textcolor{darkgreen}{0.577***(0.003) }             & \textcolor{darkgreen}{0.458***(0.003)    }          &                               \\
Size\_logsquare                       & \textcolor{red}{-0.024***(0.001) }            &\textcolor{darkgreen}{ 0.04***(0.001)    }           &                               \\
    \hline
Log-Lik.                              & -42305   & -41745   & -3.09E+05 \\
\# Obs                                & 250,000  & 250,000  & 250,000   \\
R-squared                             & 0.699   & 0.844   & 0.282     \\
F-statistic                           & 2.32E+04 & 5.41E+04 &4275    \\

\end{tabular}
\end{center}

\end{table}

\paragraph{\bf Topic of subreddit is linked to discussion tree structures:}
% \begin{itemize}
    % \item {\bf Topic of Subreddit:}
The exploratory analysis showcased in Figure~\ref{fig:grid}(a) suggests that the themes or topics of subreddits have an effect on the discussion tree structure. Topics related to {\it sports, politics, business,} and {\it local information seeking} are the ones that have more in-depth and broader discussions compared to an average discussion tree. These findings are further corroborated by the regression analysis, which controls for other variables. As can be seen from Table~\ref{reg:regresstable}, the effect sizes of all the topic features are statistically significantly different from zero. {\it Local} discussions and the ones centered around {\it sports} have the largest positive impact on all the structural properties of the discussion trees. These results are not entirely surprising as one would expect a wider variety of discussions and more intense conversations on topics related to sports and information-seeking due to the nature of these topics.

While these four types of aforementioned subreddits all typically have larger discussion tree sizes, they differ in their structure. \textit{Politics} and \textit{local information seeking} subreddits tend to be deeper, whereas \textit{sports} and \textit{business} subreddits tend to be wider. One reasoning, echoed by \citet{backstrom2013characterizing}, is that users might more often engage in back-and-forth discussions or even debates in more controversial or discussion-focused platforms of \textit{politics} and \textit{local information seeking} subreddits. In contrast, posts in \textit{sports} and \textit{business} attract a wider and more expansionary set of commentators who do not engage in further discussions.

On the other hand, subreddits related to {\it lifestyle, health}, and {\it university} seem to have sparser conversations on Reddit compared to other topics (e.g., there is about a 30\% difference compared to their counterparts in the effect on the depth of the discussion tree). However, this trend nearly vanishes after controlling for various tree properties. Some potential explanations for these narrower and shorter discussion trees could be that these topics' tend to be less controversial or that Reddit may not be the leading portal for discussions related to these topics.

\paragraph{\bf Community-level features predict discussion tree structure:}
We find that some characteristics of subreddits--redundancy, quality, novelty, and popularity--are significant predictors of discussion tree structures. Somewhat unsurprisingly, we observe that popular subreddits not only have more discussions, but those discussions are also broader. However, popular subreddits usually have fewer back-and-forth exchanges leading to deeper tree structures. In addition, our results indicate that newer subreddits have deeper discussion trees but gather a smaller set of responses overall. Perhaps older subreddits, owing to their age, have had more time to acquire subscribers to contribute to more light responses. Our results also show that subreddits containing redundant (and thus less unique information) usually have lower volume and depth of discussions but relatively {\it wide} exchanges. This is perhaps because subreddits with redundant information are also highly generic and mostly contain mundane posts that do not motivate participants to have extended conversations.

\paragraph{\bf Negative sentiment is associated with bigger and deeper trees:}
% \item {\bf Sentiment of Post:}
The model-free evidence in Figure~\ref{fig:grid} suggests that posts with weakly negative sentiment have deeper and wider discussion trees. The regression analysis (Table \ref{reg:regresstable}), which holds other variables constant, also indicates that posts with negative sentiment have higher width, depth, and quantity of conversations. One plausible explanation for this finding could be related to the contentious nature of posts with negative sentiment. Such posts often evoke strong emotions and reactions among participants, leading to heightened engagement and intense participation. Negative sentiment might provoke passionate responses, disagreements, and debates, consequently leading to extended and more expansive discussions.

\paragraph{\bf More text is associated with bigger and deeper trees:}
% \item {\bf Medium of Post:}
A Reddit post is, first and foremost, a piece of textual content a user submits to be displayed publicly online. Understanding the impact of the post's textual content can provide valuable insights into the structure of ensuing discussions. 
% The choice between different mediums, such as text or video content, can significantly influence the nature and depth of engagement a post receives.
% Thus, it is essential to quantify the impact of the post's medium on the structure of the discussions that follow. For instance, users might be interested to know whether text or video content is more likely to get higher engagement? 
Both our model-free and regression results suggest
% , perhaps, a little counter-intuitively, 
that posts with textual content, especially longer textual content, attract more broad-spectrum and high-volume discussions. The effect size of including text in the post is among the highest in terms of its positive impact on all the discussion tree's structural properties. For example, using video in a post is associated with a 53\% increase in the size of the discussion tree. We also find the length of a post's title is positively associated with the depth and size of the discussion trees. 
% One interpretation of this finding could be that a long title is more informative, thereby attracting more participants who are willing to involve in the discussion. % can attribute this finding to the fact that textual posts often lend themselves well to fostering thoughtful and extensive discussions. They allow other users to express their opinions, ideas, and arguments in a more detailed and nuanced manner, facilitating back-and-forth exchanges. Additionally, text posts often act as a catalyst for conversation starters. Users might use text to ask questions, seek advice, or share personal experiences, encouraging others to join the discussion and offer their insights. This is further supported by the significant positive effect size of the length of the posts. Longer textual posts could have the ability to provide detailed content, encourage in-depth responses, present multiple perspectives, and act as natural conversation starters.
% is more likely to stir up a thoughtful discussion garnering both back-and-forth and a broader set of participation from the audience. 
One possibility is that textual content attracts a broader audience due to its accessibility. Not all users may have the time or ability to watch a video, especially if they are browsing Reddit in public or quiet settings. Textual posts, on the other hand, can be quickly scanned, making it easier for users to participate in discussions regardless of their environment or circumstances.

\paragraph{\bf Multimedia in post is associated with wider and bigger trees:}
We observe that including a video in a post is positively associated with a significant increase in the size and width of the discussion trees but a decrease in the depth of the tree (all $p < 0.05$). Using video in a post is associated with a 5\% increase in the width of the discussion tree. One plausible explanation is that video and other multimedia content often have a high initial appeal, attracting numerous viewers due to their engaging and visually stimulating nature. However, they might not necessarily elicit as many follow-up discussions as textual posts do. Videos may convey information in a more concise and visually compelling manner, leaving less room for extended discussions or detailed responses.
    
% \item {\bf Other Observations:}

% We can make several other key observations from the results of the regression analysis. 

%\paragraph{Adult Content Lead to Larger Trees} 
%Finally, we note that posts that have adult content have discussion trees that are larger in depth, width, and size. We propose several possible reasons. Firstly, such content often involves topics that evoke strong emotions or curiosity, making it inherently attention-grabbing and engaging for users. Secondly, due to its explicit nature, NSFW content can be seen as taboo or controversial, prompting users to share diverse perspectives and opinions in response. Additionally, the shield of anonymity on Reddit allows users to freely express their thoughts and feelings, contributing to the increased volume of responses. All of these factors could lead to a deeper and wider discussion tree.

% As far as the local features are concerned, our results indicate that posts containing adult content have trees with larger depth, width, and fewer comments than the average posts. One explanation is that people feel less comfortable discussing adult content in public, especially non-adult Subreddits, leading to fewer participants engaging with such posts. 

% \end{itemize}

\begin{table*}[t]
\footnotesize
\caption{Summmary of the impact of Reddit's Global and Local Features on the Depth, Width, and Size of Discussion Trees.}

\begin{tabular}{lllllllll}
\toprule
     & \textbf{Post} & \textbf{Include} &  & && \textbf{Subreddit} & \textbf{Subreddit} & \textbf{New}\\
     & \textbf{Quality} & \textbf{Test} & \textbf{Sentiment} & \textbf{Length} & \textbf{Redundancy} & \textbf{Quality} & \textbf{Popularity} & \textbf{Subreddit}\\
\midrule
\textbf{Depth} & \textcolor{red}{Decr.}             & \textcolor{red}{Decr.}           &  \textcolor{red}{Decr.}            & \textcolor{darkgreen}{Incr.}             & \textcolor{red}{Decr.}          & \textcolor{red}{Decr.}     & \textcolor{red}{Decr.}          & \textcolor{darkgreen}{Incr.}         \\
\textbf{Width} & \textcolor{darkgreen}{Incr.}             & \textcolor{darkgreen}{Incr.}            & \textcolor{darkgreen}{Incr.}            & \textcolor{red}{Decr.}            & \textcolor{darkgreen}{Incr.}            & \textcolor{darkgreen}{Incr.}      & \textcolor{darkgreen}{Incr.}          & \textcolor{red}{Decr.}          \\
\textbf{Size}    & \textcolor{darkgreen}{Incr.}             & \textcolor{darkgreen}{Incr.}            & \textcolor{red}{Decr.}              &  \textcolor{darkgreen}{Incr.}           & \textcolor{red}{Decr.}         & \textcolor{red}{Decr.}      & \textcolor{red}{Decr.}          & \textcolor{red}{Decr.}        
\\

\end{tabular}

\label{tab:summary}
\end{table*}

\section{Discussion}
Threaded online discussions are ubiquitous across various social media platforms, inducing graphical tree structures. Our research explored three vital structural aspects of these discussion trees: width, depth, and size, through an empirical analysis of over 250,000 posts and 3 million comments. We hypothesized that the structure of online discussions depends on both the broader environment (global features) and the specific properties of the content shared (local features). To test this hypothesis, we examined the impact of a comprehensive set of global features (representing the subreddit properties) and local features (representing post-specific attributes).

Our findings reveal that while both global and local features are crucial in explaining discussion trees' structural properties, local features collectively explain significantly more variation than global features. This suggests that users primarily decide to interact with a post based on its specific attributes rather than the characteristics of the subreddit. Nevertheless, several global properties of subreddits also significantly influence discussion tree structures, emphasizing the need to consider both feature types.

Previous social media studies have emphasized the importance of local features, such as popularity and novelty, in determining post spread~\cite{gomez2008statistical,lumbreras2016automatic,aragon2017thread}. Our study addresses a gap in the literature by considering both local and global features, drawing inspiration from related work exploring how these features influence various aspects of online communication~\cite{zhang2017community,lambert2022conversational,coletto2017automatic}. By quantifying the relative importance of global versus local features in shaping discussion tree structures, we provide a more comprehensive understanding of the factors driving information spread on social media platforms.

Overall, our study sheds light on the significance of local and global features in shaping discussion structures and provides practical design implications for platforms that facilitate curated discussions. Furthermore, our findings offer insights into developing discussion predictors that can assist content moderators in effectively identifying and addressing potential issues.

\subsection{Impact of Global Features}
Our research highlights the crucial role subreddits play in determining discussion structure. As hubs that provide the substrate for conversations, subreddits define community culture and moderation guidelines, indirectly impacting users' propensity to interact with or create posts. We found that specific subreddit properties, such as topic and information redundancy, contain significant explanatory power. For instance, subreddits discussing news, politics, or information-seeking topics tend to have deeper and broader discussion trees with many back-and-forth conversations. This aligns with previous work by~\citet{gonzalez2010structure} on political topics, which found that politics-related issues tend to have higher participation rates and longer chains of information exchange than non-political topics. The controversial nature of such content, coupled with its general applicability, likely leads to high engagement. Similarly, sport-related subreddits exhibit substantial engagement, leading to extensive and profound discussions. This can be attributed to the intense emotional connection and strong sense of group identification that sports fans tend to foster with their favorite teams, compared to other fan communities~\cite{malin2005energy}. In contrast, ``lifestyle’’ content generates comparatively leaner discussion trees, possibly due to lower levels of emotional attachment and group identification among users.

\subsection{Impact of Local Features}
Our analysis revealed several important local features that predict the size, depth, and width of discussion trees. A key finding is that posts with negative sentiment attract both broad and in-depth conversation, supporting previous work by~\citet{mejova2014controversy} on the framing of controversial issues in news. Conversely, posts expressing neutral emotions tend to have fewer follow-up discussions. This suggests that emotionally charged content and posts with high arousal involve more participants and generate more back-and-forth conversations, aligning with research on the potential of emotionally charged tweets to be retweeted~\cite{stieglitz2013emotions} and the ability of high-valence news to spread wider~\cite{berger2012makes}.

Another notable finding is the ability of textual content to generate broader discussions and attract high participation, while video content tends to spread widely but does not stimulate in-depth follow-up discussions. This difference may be because text-based posts encourage audience engagement with the original poster's ideas, leading to back-and-forth conversations, whereas multimedia content might attract attention but fail to provoke deeper discussion.

Our results can help researchers further refine their understanding of conversational success and discussion norms. Previous studies have identified time-related variables, such as early conversation content, politeness, and rhetorical prompts as essential features in detecting conversation success~\cite{zhang2018conversations,hessel2019something}. Our findings suggest that local features are substantially more predictive than global features in such prediction tasks.

% \subsection{Design implications for discussion curation in social media platforms }
\subsection{Implications}
This research contributes significantly to the broader CSCW research community by offering valuable insights into online conversation dynamics across different community groups. Our findings have practical implications for both users and platform designers in improving the user experience on discussion forums like Reddit.

\subsubsection{User Implications}
For content producers, such as news organizations or digital marketing agencies, our findings suggest a strategic approach to stimulating meaningful online discussions. The emphasis should be on optimizing the properties of the content itself (local features) rather than solely focusing on selecting the ``correct’’ environment (global features) for dissemination. Content creators should prioritize producing high-quality, emotionally engaging content that incorporates both textual and video elements. Focusing on less redundant environments can increase visibility, while using textual elements encourages in-depth, back-and-forth discussions. Video content can be utilized for broader reach, albeit with potentially less focused discussion. While publishing on popular platforms can increase visibility, it should be a secondary consideration due to the risk of content being crowded out. By following these guidelines, content creators can optimize their posts for maximum engagement and meaningful discussion.

\subsubsection{Platform Design Implications}
Our study underscores the importance of both local and global features in shaping discussion structures, with a stronger emphasis on local features. However, global features, particularly topic and subreddit redundancy, also play significant roles. Based on these findings, we propose two key design considerations for platforms: offering subreddit recommendations for end-users and providing content alerts for subreddit moderators.

\begin{itemize}

\item {\bf Offering subreddit recommendations for end-users:}
Platform designers should leverage the heterogeneity of global and local features when developing algorithms to support and enhance discussions. Notably, specific global features, such as the topic, age, popularity, and content redundancy in a subreddit, have been identified as influential factors in shaping the characteristics of discussion trees. Platforms can leverage these insights to inform their design strategies, including offering recommendations for relevant subreddits, promoting engaging topics, and addressing redundancy concerns within discussions. By incorporating these design considerations, platforms can foster more effective and meaningful conversations among users. Additionally, platform designers could implement a one-time prompt to assist users interested in promoting meaningful discussions. For instance, the prompt could provide suggestions on the tone (sentiment) used to convey information or recommend adding more context to encourage engagement in the discussion. This proactive approach can help users improve their post quality and select appropriate subreddits. By incorporating these design elements, platforms can foster more effective and meaningful conversations among users, enhancing overall user experience and community engagement.

\item {\bf Providing content alert for subreddit moderators:}
Our findings offer valuable insights for preventing collective violence within discussions, a critical issue given the scarcity of human moderators on online platforms. Subreddit moderators are individuals who volunteer their time to oversee, guide, and nurture a subreddit community, assuming responsibility for setting its tone, moderating content, and enforcing user bans when necessary. These moderators operate independently of Reddit as a company, dedicating themselves to serving the community's best interests. 
 Subreddit moderators can benefit from tools that combine lexical cues (e.g., politeness~\cite{zhang2018conversations}) with our discussion structure predictors to identify potentially problematic posts early. These tools can help moderators prioritize their efforts, intervene proactively in potentially problematic discussions, and maintain a healthier community environment.
 \end{itemize}

\subsection{Limitations}
While our study provides valuable insights, it's important to acknowledge its limitations. The study focuses solely on Reddit, potentially limiting its applicability to other platforms. Our study design allows us to identify correlations but not establish causal relationships between features and discussion structures. We rely on Pushshift for data, which may contain non-systematic gaps and deleted information~\cite{gaffney2018caveat}, although our 2018 dataset shows uniform distribution across time. The structure of discussion trees may be influenced by UI design choices, which our study doesn't address due to lack of usage pattern data. Some features show smaller effect sizes on discussion structure, possibly due to factors like user culture and content-specific meanings. Finally, online discussions serve various purposes, each potentially resulting in different thread structures. Our study aims to elucidate how content and community context contribute to these variations rather than promoting any particular type of discussion tree. These limitations highlight opportunities for future research, including cross-platform studies, causal analyses, and investigations into the impact of UI design on discussion structures.

\section{Conclusion}
This paper examined the structure of threaded discussions on social media, focusing on the key structural attributes of discussion posts: width, depth, and size. By analyzing both local and global discussion tree properties, we identified the factors that explain why certain discussion threads evolve into deep, thoughtful conversations, while others generate a wide array of responses, and some achieve both depth and breadth. Our work makes key contributions, including a comprehensive analysis of the interplay between local features (specific to individual posts) and global features (characteristics of the broader community or platform), and a novel framework for categorizing and analyzing discussion trees. Another significant contribution of our work is the creation of a new large-scale dataset that captured the complex dynamics of online discussion structures.

Finally, our findings have far-reaching implications for both content creators and platform designers. For content creators, our results offer insights into crafting posts that are more likely to generate engaging and meaningful discussions. Platform designers could leverage our findings to develop more effective recommendation systems, moderation tools, and user interfaces that encourage productive discussions. By unpacking the anatomy of threaded discussions, our work offered both theoretical insights and practical tools for shaping more constructive and engaging online spaces. As digital communication continues to evolve, the frameworks and findings presented in this paper will serve as valuable resources for researchers, platform developers, and policymakers working to improve the quality of online discourse.

\bibliographystyle{ACM-Reference-Format}
\bibliography{cite}

\section{Appendix}
Here's the list of all the 500 subreddits used in this paper:
\noindent 
{\footnotesize
`USLPRO', `internetparents', `onewordeach', `6thForm',
       `breakingmom', `AFROTC', `DotA2', `gachagaming', `fiaustralia',
       `Eugene', `cakeday', `navy', `SteamController',
       `AmericanHorrorStory', `podcasting', `IncreasinglyVerbose',
       `Roadcam', `entwives', `anchorage', `greece', `DianaMains',
       `PokelandLegends', `ShrugLifeSyndicate', `Documentaries',
       `TwoBestFriendsPlay', `WC3', `creepyencounters', `MLPLounge',
       `SuperMarioOdyssey', `TheFrontBottoms', `pathofdiablo', `AskMen',
       `gzcl', `WhiteWolfRPG', `Otonokizaka', `modeltrains', `eurovision',
       `Boardgamedeals', `insanepeoplefacebook', `Militaryfaq', `Posture',
       `personalfinance', `pocketoperators', `ufl', `GooglePixel',
       `LifeProTips', `AjaxAmsterdam', `angelsbaseball',
       `wildhearthstone', `G502MasterRace', `MysteryDungeon',
       `Stronglifts5x5', `redsox', `PS4', `rugbyunion', `TrueChristian',
       `PHPhelp', `Waluigi', `MacOS', `KitchenConfidential',
       `TwoXChromosomes', `PartyParrot', `Re\_Zero', `Neverwinter',
       `Triumph', `GuitarAmps', `LDESurvival', `guitarpedals',
       `Anticonsumption', `crash\_fever', `ussoccer', `quityourbullshit',
       `skyrim', `askgaybros', `piercing', `progresspics', `iOSthemes',
       `headphones', `MurderedByWords', `SpaceBuckets',
       `allthingsprotoss', `FLMedicalTrees', `Gloomhaven', `Cartalk',
      `DnDBehindTheScreen', `AskGames', `AskProgramming',
      `WarhammerUnderworlds', `AnimalCrossingNewLeaf', `Pokemongiveaway',
      `deadcells', `latterdaysaints', `Bonsai', `bodyweightfitness',
      `PokemonPlaza', `Breadit', `menstrualcups', `Libraries',
      `ballpython', `dayton', `Porsche', `AnxietyDepression', `EDC',
      `canadients', `bulletjournal', `trashy', `njpw', `darkestdungeon',
      `SuicideWatch', `ultimate', `nova', `phillies', `Drama',
      `RandomActsofMakeup', `NevilleGoddard', `archlinux', `deftones',
      `UofT', `IsaacArthur', `nationalguard', `betterCallSaul',
       `twinpeaks', `dreamcatcher', `mixer', `fitnesscirclejerk',
       `HustleCastle', `ClashRoyale', `rulesofsurvival', `HollowKnight',
       `FinancialCareers', `sadboys', `VitaPiracy', `manga', `mvci',
       `TargetedShirts', `travisscott', `tarot', `Alienware',
       `R6ProLeague', `EngineeringPorn', `socalhiking', `migraine',
       `askMRP', `satanism', `buffy', `grilledcheese', `greenville',
       `IncrediblesMemes', `NoMansSkyTheGame', `viktormains', `amiibo',
       `Frugal', `bingingwithbabish', `ElgatoGaming', `salesforce',
       `Narcolepsy', `RDR2', `Psychonaut', `sportsbook',
       `SkincareAddiction', `botany', `rape', `MUAontheCheap', `RimWorld',
       `Trove', `MLBTheShow', `electricdaisycarnival', `BreakingBenjamin',
       `FreeKarma4You', `ftm', `CarAV', `Charcuterie', `Planetside',
       `metacanada', `seduction', `TheCannalysts', `stockholm',
       `assassinscreed', `3DS', `getting\_over\_it', `maschine',
       `Surveying', `birthcontrol', `FishMTG', `Detroit',
       `TooAfraidToAsk', `Portal', `lawschooladmissions', `pixel2',
       `Cash4Cash', `bobdylan', `LPOTL', `MaddenUltimateTeam', `XWingTMG',
       `pkmntcgcollections', `BlackClover', `miband', `policydebate',
       `DrDisrespectLive', `maybemaybemaybe', `synology', `LoveIslandTV',
       `prusa3d', `fantasybaseball', `rheumatoid', `ComedyCemetery',
       `ptsd', `blackberry', `milwaukee', `SavageGarden',
       `badwomensanatomy', `arrow', `southpark', `cpp', `DisneyPinSwap',
       `Ducati', `justlegbeardthings', `microdosing',
       `AdventuresOfSabrina', `instantpot', `UGA', `TheDarkTower',
       `brandonsanderson', `RecRoom', `AtlantaUnited',
       `miraculousladybug', `1P\_LSD', `3d6', `keto', `lasercutting',
       `RedvsBlue', `Polandballart', `overclocking', `moomooio',
  `gameofthrones', `hoggit', `riverdale', `Adelaide',
  `PlanetCoaster', `haikyuu', `casualiama', `criticalrole',
  `GalaxyNote8', `gundeals', `TrueDoTA2', `ffxiv', `olympia',
  `funkoswap', `Schizoid', `uwaterloo', `PacificCrestTrail',
  `cinematography', `ThailandTourism', `Guitar', `Miata', `YIMO',
  `Screenwriting', `magicTCG', `ShokugekiNoSoma', `spicy',
  `ModestMouse', `jakeandamir', `ExplainAFilmPlotBadly', `askdrugs',
  `Paranormal', `LARP', `totalwar', `PhillyUnion', `conlangs',
  `freefolk', `BossfightUniverse', `Justfuckmyshitup', `askTO',
  `civ', `BattleRite', `savannah', `BigBrother', `DarK', `Swimming',
  `RatchetAndClank', `Gintama', `LearnCSGO', `projectzomboid',
  `PoliticalVideo', `hulaween', `zootopia', `wisconsin',
  `bioniclememes', `rollercoasters', `xxfitness', `OrderOfHeroes',
  `KansasCityChiefs', `essential', `WWEChampions', `jerseyshore',
  `NoContract', `UberEATS', `chinaglass', `bindingofisaac',
  `kemonomimi', `airsoftmarket', `progun', `OpiatesRecovery',
  `ps4homebrew', `firefly', `linux4noobs', `Brogress', `Polytopia',
  `energydrinks', `oddlysatisfying', `forbiddensnacks', `SCUMgame',
  `SSBPM', `SelfHarmScars', `jetta', `Winnipeg', `TokyoGhoul',
  `COMPLETEANARCHY', `JurassicPark', `vegas', `NolanBatmanMemes',
  `Omaha', `oakland', `wichita', `Reno', `BloodAngels',
  `Aurelion\_Sol\_mains', `toarumajutsunoindex', `uBlockOrigin',
  `InfinityTheGame', `whatisthisthing', `TeenAmIUgly', `SS13',
  `Waxpen', `CatAdvice', `melbourne', `oklahoma', `PMDD',
  `panelshow', `uvtrade', `origami', `Citra', `survivor',
  `ImaginaryLeviathans', `MakeupAddiction', `InternetStars',
  `Konosuba', `socialwork', `homegym', `7daystodie', `KhaZixMains',
  `TheStrokes', `GEazy', `SeattleWA', `amathenedit', `shameless',
  `forhonorknights', `CurseofStrahd', `DarkSouls2', `OLED',
  `Heroclix', `fordranger', `fatestaynight', `VintageApple',
  `whatsthisbird', `yakuzagames', `Unity3D', `mechmarket',
  `GoneMildPlus', `penpals', `darknet', `IBO', `Destiny6Global',
  `exchangeserver', `Wellington', `PersonalFinanceCanada',
  `lisathepainfulrpg', `DataHoarder', `blackmagicfuckery',
  `EliteDangerous', `ar15', `bikesgonewild', `guitarcirclejerk',
  `Albany', `FL\_Studio', `EnaiRim', `delusionalartists', `MephHeads',
  `GamePhysics', `iastate', `NetflixBestOf', `Magicdeckbuilding',
  `ANormalDayInRussia', `BeautyBoxes', `newfoundland', `ZeroWaste',
  `volt', `creepyasterisks', `UKPersonalFinance', `BattlefieldV',
  `hardwareswap', `Smite', `Ghostbc', `picrequests', `Tendies',
  `ProRevenge', `CharlotteHornets', `climbharder', `stepparents',
  `WeAreTheMusicMakers', `Rateme', `CanadaPolitics', `DuelLinks',
  `OnePlus6', `shield', `vaginismus', `TheBluePill', `wokekids',
  `USMC', `SuddenlyGay', `ccna', `whichbike', `cscareerquestions',
  `Persona5', `PSVR', `SneakerFits', `FireflyFestival', `masseffect',
  `BendyAndTheInkMachine', `Dirtybomb', `lolgrindr', `grandrapids',
  `Discord\_Bots', `toronto', `Blacksmith', `crazyexgirlfriend',
  `cigars', `vegan', `dating\_advice', `addiction', `skyrimrequiem',
  `DiWHY', `uncharted', `swdestiny', `GalaxyS8', `AskMenAdvice',
  `Boruto', `cfs', `talesoftherays', `schizophrenia',
  `queensuniversity', `ecobee', `cactus', `makeupflatlays',
  `AskScienceFiction', `MushroomGrowers', `iamverybadass', `kol',
  `linux\_gaming', `customhearthstone', `Prematurecelebration',
  `phoenix', `RetroFuturism', `Supernatural', `forza', `CodingHelp',
  `CruciblePlaybook', `churning', `BanPitBulls', `spaceporn',
  `Mariners', `lgg7', `SHINee', `engineering', `TaliyahMains',
  `MaladaptiveDreaming', `ECEProfessionals', `Calgary',
  `thalassophobia', `BeardedDragons', `nyu', `Jeep', `HotlineMiami',
  `Guiltygear', `whatcarshouldIbuy', `MtvChallenge', `mobilelegends',
  `Socialism\_101', `vegancirclejerk', `git', `overlanding',
  `Concordia', `django', `freemasonry', `Megaman', `makeupexchange',
  `weightlifting', `XVcrosstrek', `shortcuts', and `disneyemojiblitz'.}

\end{document}